\documentclass[fleqn,usenatbib]{mnras}

\usepackage{newtxtext,newtxmath}

\usepackage[T1]{fontenc}

\DeclareRobustCommand{\VAN}[3]{#2}
\let\VANthebibliography\thebibliography
\def\thebibliography{\DeclareRobustCommand{\VAN}[3]{##3}\VANthebibliography}

\usepackage{graphicx}	
\usepackage{amsmath}	
\usepackage{xcolor}     

\title{Density waves in protoplanetary discs excited by eccentric planets: linear theory}

\author[Fairbairn \& Rafikov]{
Callum W. Fairbairn$^{1}$\thanks{E-mail: cwf29@cam.ac.uk}
and Roman R. Rafikov$^{1,2}$\thanks{E-mail: rrr@damtp.cam.ac.uk}
\\
$^{1}$Department of Applied Mathematics and Theoretical Physics, CMS, University of Cambridge, Wilberforce Road, Cambridge CB3 0WA, UK \\
$^{2}$Institute of Advanced Study, Einstein Drive, Princeton, NJ 08540, USA
}

\date{Accepted Year Month Day. Received Year Month Day; in original form Year Month Day}

\pubyear{2022}

\begin{document}
\label{firstpage}
\pagerange{\pageref{firstpage}--\pageref{lastpage}}
\maketitle

\begin{abstract}
Spiral density waves observed in protoplanetary discs have often been used to infer the presence of embedded planets. This inference relies both on simulations as well as the linear theory of planet-disc interaction developed for planets on circular orbits to predict the morphology of the density wake. In this work we develop and implement a linear framework for calculating the structure of the density wave in a gaseous disc driven by an eccentric planet. Our approach takes into account both the essential azimuthal and temporal periodicities of the problem, allowing us to treat any periodic perturbing potential (i.e. not only that of an eccentric planet). We test our framework by calculating the morphology of the density waves excited by an eccentric, low-mass planet embedded in a globally isothermal disc and compare our results to the recent direct numerical simulations (and heuristic wavelet analysis) of the same problem by Zhu and Zhang. We find excellent agreement with the numerical simulations, capturing all the complex eccentric features including spiral bifurcations, wave crossings and planet-wave detachments, with improved accuracy and detail compared with the wavelet method. This illustrates the power of our linear framework in reproducing the morphology of complicated time-dependent density wakes, presenting it as a valuable tool for future studies of eccentric planet-disc interactions.
\end{abstract}









\section{Introduction}
\label{section:introduction}

Many protoplanetary discs are known to exhibit a rich variety of substructures \citep{Andrews2020}, with large scale spiral arms being one of the most impressive examples of the departures from axisymmetry. Direct imaging of protoplanetary discs in the near-IR has shown the ubiquity of spirals in a variety of  systems including MWC 758 \citep[e.g.][]{GradyEtAl2012,BenistyEtAl2015,ReggianiEtAl2018}, HD100453 \citep[e.g.][]{WagnerEtAl2015,WagnerEtAl2018}, SAO 206462 \citep[e.g.][]{MutoEtAl2012} and LKHa 330 \citep{UyamaEtAl2018}. High-resolution observations of protoplanetary discs in sub-mm by the \textit{Atacama Large Millimetre Array} (ALMA) have also illuminated many spiral features. In particular, spirals have been found in dust continuum emission in Elias 2-27, IM Lup, and WaOph 6 \citep{PerezEtAl2016,HuangEtAl2018}, whereas CO line emission has revealed spirals in the gaseous component of the protoplanetary disc in AB Aurigae \citep{TangEtAl2017}. 

One popular hypothesis for explaining spiral features is through the existence of planets embedded within the disc, which interact gravitationally with the surrounding material launching density waves. The linear theory of such planet-disc interaction was developed by \citet{GoldreichTremaine1979,GoldreichTremaine1980} and \citet{LP1979} as motivated by the galactic spiral density wave theory \citep{LinShu1964}. According to this theory, a massive perturber (a planet) provides a periodic forcing to the fluid elements in the disc which resonates with the epicyclic frequency at a set of Lindblad resonances. The gas pressure (and self-gravity in more massive discs) then communicates this forcing through the disc in the form of a density wave, which appears as a spiral structure due to the differential rotation of the underlying disc. 

Historically, planet-disc interaction theory has been used primarily to study the excitation of density waves and calculate the amount of angular momentum they carry. Understanding the deposition of this angular momentum in the disc is a subject of key importance for planet migration and gap opening. However, in recent years, partly motivated by the observations of spiral structures in protoplanetary discs \citep[e.g.][]{AvenhausEtAl2014,BenistyEtAl2015,MonnierEtAl2019}, the mathematical apparatus of the linear planet-disc interaction theory has also been employed to understand the morphology of the planet-driven density wake, and to use this feature to infer the properties of embedded protoplanets. 

In particular, the linear analysis performed by \cite{OgilvieLubow2002} showed that a low mass planet launches a one-armed (primary) spiral arm in the disc, which results from the constructive superposition of many wave modes. Subsequent studies have shown that this situation is, in fact, more complex, and that {\it multiple} spiral arms can be caused by a single planet (at least in the inner disc). This phenomenon was first found in fully-nonlinear hydrodynamical simulations of protoplanetary discs \citep{DongEtAl2015,ZhuEtAl2015,DongFung2017}, but later \citet{Bae2018} and  \cite{MirandaRafikov2019} demonstrated that such higher-order spirals are an intrinsically linear phenomenon resulting from the dispersive nature of the density waves. Subsequently, \cite{MirandaRafikov2020} and \cite{MirandaRafikov2020b} have shown that incorporating additional physics (e.g. gas cooling) can significantly affect wave propagation and increase the diversity of spiral morphologies. However, quite importantly, all such morphological studies have focused on the case of a planet moving on a {\it circular} orbit.

In recent years, there emerged a number of observations which cannot be easily explained in the framework of a zero-eccentricity (circular) planet-disc interaction. For example, the strong variation observed in pitch angles of spirals  \citep[e.g.][]{UyamaEtAl2020} is unlikely to be caused by low mass planets on circular orbits, which tend to have shallow pitch angles away from the planet. Furthermore, there is tentative evidence of time variability within the two spiral arms of SAO 206462, which might exhibit different pattern speeds \citep{XieEtAl2021}. These effects might be explained by a planet on an {\it eccentric} orbit which launches an intrinsically time-dependent wake. This possibility is not at all exotic since a number of exoplanets are found to have eccentric orbits \citep[e.g.][]{KaneEtAl2012,XieEtAl2016,EylenEtAl2019}. Several authors have explored the possibility that these eccentricities might arise due to planet-planet resonances or scattering events \citep{FordRasio2008,ChatterjeeEtAl2008,PetrovichEtAl2014}, by means of gravitational interaction with the nascent disc  \citep{GoldreichSari2003,TeyssandierOgilvie2016a,TeyssandierOgilvie2017,RagusaEtAl2018}, or due to a thermal torque maintaining high eccentricities for low mass planets \citep{Masset2017,EklundMasset2017,FromenteauMasset2019}.

Recently \cite{ZhuZhang2022} (hereafter ZZ22) have carried out a set of hydrodynamic simulations of eccentric planets embedded in discs, finding a rich variety of the density wave morphologies emerging as a functions of the planetary eccentricity $e$: crossings of the spiral arms, sharp changes in pitch angles, bifurcating spirals, and so on. They also came up with a simple heuristic model (a Huygens-like wavelet analysis) for describing these wake morphologies, which compares rather favourably with their simulations, while missing out on some of the pattern complexity and amplitude information. 

In this work we develop a fully time-dependent, linear, semi-analytical framework for exploring the morphology of the density waves launched by a time-periodic perturbing potential, and apply it to the case of a planet on an eccentric orbit. Being linear by construction, our framework naturally complements the fully non-linear simulations of ZZ22, allowing one to cleanly separate the role of linear effects in setting the planet-driven wave pattern. On the other hand, our approach presents a fast and efficient method for exploring the morphology of planetary wakes, while also fully preserving the wave amplitude information. 

The paper is organised as follows. In  \S\ref{section:framework} we outline our general setup and develop the mathematical apparatus based on the linear perturbation analysis, with the master equation (\ref{eq:master_eqn}) being our key result. In \S\ref{section:numerical} we describe the numerical techniques used to solve the master equation and construct the pattern of the spiral structure driven by an eccentric planet. In \S\ref{section:results} we examine the wake morphologies predicted by our framework for a range of planetary eccentricities and compare them with the numerical simulations/wavelet approach of ZZ22. Finally we discuss the advantages of our method in  \S\ref{section:discussion} and summarise in  \S\ref{section:conclusions}.
\section{Problem framework}
\label{section:framework}

\subsection{Planet disc configuration}

\begin{figure}
    \centering
    \includegraphics[width=0.8\columnwidth]{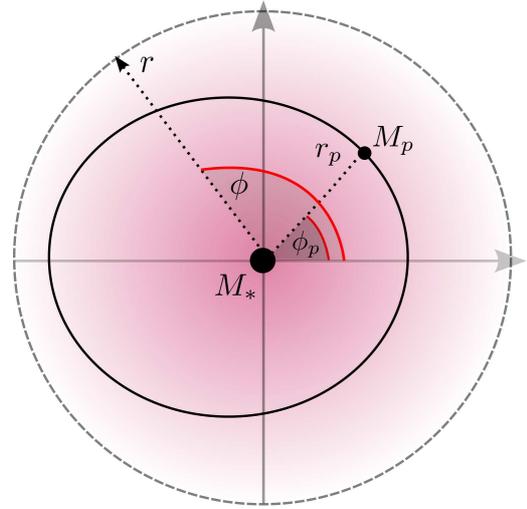}
    \caption{Schematic illustration of the geometry of the problem. The central star has mass $M_*$ and is at the origin of the polar coordinate grid. The eccentric orbit of the planet is indicated as the solid black line with the planet marked with polar coordinates $(r_{\text{p}},\phi_{\text{p}})$. An arbitrary point in the disc is then denoted $(r,\phi)$. }
    \label{fig:cartoon}
\end{figure}

We will consider a planet of mass $M_{\text{p}}$ moving on a fixed eccentric orbit about the host star of mass $M_*$ as shown in Fig.~\ref{fig:cartoon}.
The planetary orbit is parameterised by the eccentricity $e$ and semi-major axis $a$. Planetary mean motion $n$ is equal to the Keplerian rotation rate $\Omega_K = (G M_*/r^3)^{1/2}$ evaluated at the planetary semi-major axis $r=a$. We adopt a reference frame centred on the primary star with position labelled by the radial and azimuthal coordinates $(r,\phi)$. In this frame the gravitational acceleration due to the perturbing planet is given by the disturbing function \citep{MurrayMcDermott1999},
\begin{equation}
    \label{eq:disturbing_function}
    \mathcal{R}(\mathbf{r}) = -\Phi-G M_{\text{p}}\frac{\mathbf{r}_{\text{p}}\cdot \mathbf{r}}{r_{\text{p}}^3},
\end{equation}
where
\begin{equation}
    \label{eq:direct_potential}
    \Phi(\mathbf{r}) = -\frac{G M_{\text{p}}}{|\mathbf{r}-\mathbf{r}_{\text{p}}|},
\end{equation}
is the direct potential term, $\mathbf{r}_{\text{p}}$ is the position vector of the planet and $\mathbf{r}$ is an arbitrary position vector in the disc relative to the host star. The second, indirect term in the disturbing function arises from the stellar motion around the system barycenter. Since ZZ22 neglect the indirect term in their low mass planet numerical simulations, we will do the same here for fairer comparison between results. In practice we find that inclusion of the indirect term makes little difference to net the wake morphology, as per the previous findings of \cite{MirandaRafikov2019} for the circular planet case. This will be examined further in a future paper exploring the full parameter space of our framework.

The planetary orbit is fully embedded within a (co-planar) inviscid gaseous disc, which we will consider as a two-dimensional structure. We prescribe an unperturbed, axisymmetric background disc state, according to the power law profiles for the surface density
\begin{equation}
    \label{eq:framework:sigma_background}
    \Sigma(r) = \Sigma_{\text{p}} \left(\frac{r}{a}\right)^{-p},
\end{equation}
with $\Sigma_{\text{p}}$ being the surface density at $r=a$, and for the isothermal sound speed
\begin{equation}
    \label{eq:framework:csiso_background}
    c_{s,\text{iso}}(r) = h_{\text{p}} v_K(a) \left(\frac{r}{a}\right)^{-q/2},
\end{equation}
where $v_K(a)=an$ is the Keplerian velocity of the planetary guiding centre. Meanwhile $h_{\text{p}}$ is the disc aspect ratio, $h(r) = H/r = h_{\text{p}}(r/a)^{(1-q)/2}$ evaluated at $r=a$, where $H = c_{s,\text{iso}}/\Omega_K$ is the pressure scale height. The disc temperature profile is simply proportional to $c_{s,\text{iso}}^2$ and is controlled by the exponent $q$.

The product of these quantities sets the vertically integrated pressure through the {\it locally isothermal} equation of state (EoS)
\begin{equation}
\label{eq:eos}
    P = \Sigma c_{s,\text{iso}}^2.
\end{equation}
Our linear, semi-analytical framework is developed in \S\ref{sec:pert-eq} for this rather general EoS. However, later in this work we will restrict our attention to the more restrictive case of a {\it globally isothermal} EoS with $q=0$, as this EoS is known to conserve wave angular momentum flux away from the Linblad resonances \citep{MirandaRafikov2019b}. 

This background radial structure generally introduces a pressure gradient which modifies the usual Keplerian rotation, $\Omega_K$, such that the fluid orbital frequency becomes
\begin{equation}
    \label{eq:framework:orbital_frequency}
    \Omega^2 = \Omega_K^2+\frac{1}{r\Sigma}\frac{dP}{dr}=\Omega_K^2\left[1-h_{\text{p}}^2(p+q)\left(\frac{r}{a}\right)^{1-q}\right].
\end{equation}
Thus the background azimuthal velocity profile is $u_\phi(r) = r \Omega(r)$, whilst the unperturbed radial velocity in the absence of viscosity is $u_r = 0$. The resulting radial gradient of specific angular momentum sets the epicyclic frequency $\kappa$ of the fluid given by 
\begin{equation}
\label{eq:framework:eq:epicyclic}
    \kappa^2(r) = \frac{2\Omega}{r}\frac{d}{dr}(r^2\Omega)=\Omega_K^2\left[1-h_{\text{p}}^2(p+q)(2-q)\left(\frac{r}{a}\right)^{1-q}\right],
\end{equation}
to be slightly different from $\Omega_K(r)$.

\subsection{Linearized perturbation equations}
\label{sec:pert-eq}

To study the morphology of the density waves produced by eccentric planets in protoplanetary discs we generalise the methodology described in \cite{MirandaRafikov2019}, for treating the circular version of the problem, towards eccentric orbits. This approach exploits the fundamental linear theory of planet-disc interactions originally developed in \cite{GoldreichTremaine1979,GoldreichTremaine1980}. It is well known \citep{Goodman2001} that for the planet-disc interaction to remain in the linear regime the mass of the perturbing planet must remain small compared to the so-called {\it thermal mass}
\begin{equation}
\label{eq:thermal_mass}
    M_\mathrm{th}=h_{\text{p}}^3M_* \ll M_*.
\end{equation}
Thus, throughout this study we will assume  $M_\mathrm{p}\lesssim M_\mathrm{th}$.

The key difference of the present development compared to previous work is that the planet is no longer assumed to move on a circular orbit. Thus, the time dependence of the planetary potential cannot be eliminated by a simple change of coordinates to a frame co-rotating with the planet (wherein, for the circular case, the perturbing gravitational potential is stationary and the excited wake forms a steady-state structure). In the eccentric case the planet oscillates about the guiding centre of its orbit (which is in uniform rotation around $M_*$) both azimuthally and radially, completing a closed loop\footnote{For small values of the planetary  eccentricity, the epicyclic approximation holds and the planet moves on an elliptical trajectory centred on the guiding centre with maximal azimuthal displacement equal to twice that of the radial displacement.} once per orbit. Thus, there are now two intrinsic periodicities in the problem --- azimuthal and temporal --- and we can expand the potential as a Fourier series \citep{GoldreichTremaine1980}
\begin{equation}
    \label{eq:fourier_expansion}
    \Phi = \sum_{m = 1}^{\infty} \sum_{l = -\infty}^{\infty} \mathrm{Re}\left[\Phi_{lm}(r)e^{i(m\phi-l n t)}\right],
\end{equation}
with modal numbers $l$ and $m$ and $\Phi_{lm}$ being the Fourier amplitudes of the perturbing potential further discussed in \S\ref{subsection:extracting_potential}. We are free to set the time origin such that the planet is aligned with $\phi = 0$ at $t = 0$. Then the planetary potential is even in azimuth and $\Phi_{lm}$ is real, resulting in
\begin{equation}
    \label{eq:cosine_expansion}
    \Phi = \sum_{m = 1}^{\infty} \sum_{l = -\infty}^{\infty} \Phi_{lm}(r)\cos\left(m(\phi-\omega_{lm} t)\right).
\end{equation}
Notice that each $(l,m)$ Fourier component has a pattern speed
\begin{equation}
\label{eq:olm}
\omega_{lm} = \frac{l}{m}n, 
\end{equation}
which differs from the mean motion $n$ when $l \ne m$. These forcing components will excite a strong response and launch waves in the disc at particular eccentric Linblad resonances which are nicely summarised in the work of \cite{GoldreichSari2003} who were more interested in the feedback on the planet rather than the wake morphology. 

This disturbing potential perturbs the background disc such that $\Sigma \rightarrow \Sigma +\delta \Sigma$, $P \rightarrow P +\delta P$, $u_r \rightarrow u_r +\delta u_r$ and $u_\phi \rightarrow u_\phi +\delta u_\phi$. All the perturbed quantities are assumed to have a Fourier harmonic form, forced by each mode of the planetary potential presented in equation \eqref{eq:fourier_expansion}, such that
\begin{equation}
    \delta x(r,\phi, t) = \mathrm{Re} \left[\delta x(r)e^{im(\phi-\omega_{lm}t)} \right]
\end{equation}
for a perturbed variable $x$.

The net perturbation is then the sum of each of these forced Fourier contributions. Substituting this ansatz into the linearised hydrodynamic equations for two-dimensional mass and momentum conservation we arrive at the usual equations
\begin{equation}
\label{eq:density_pert_eqn}
    -i \Tilde{\omega}\delta \Sigma + \frac{1}{r}\frac{\partial}{\partial r}(r\Sigma \delta u_r)+\frac{i m \Sigma}{r}\delta u_\phi = 0 , 
\end{equation}

\begin{equation}
\label{eq:ur_eqn}
    -i \Tilde{\omega}\delta u_r - 2\Omega \delta u_{\phi} = -\frac{1}{\Sigma}\frac{\partial}{\partial r} \delta P +\frac{1}{\Sigma^2}\frac{dP}{dr}\delta \Sigma-\frac{\partial}{\partial r}\Phi_{lm},
\end{equation}

\begin{equation}
\label{eq:uphi_eqn}
-i \Tilde{\omega}\delta u_\phi + \frac{\kappa^2}{2\Omega}\delta u_r = -\frac{i m}{r}\left( \frac{\delta P}{\Sigma}+\Phi_{lm}\right),
\end{equation}
which appear identical to equations (2)--(4) in \cite{MirandaRafikov2020}. The subtle difference is now wrapped up in the Doppler-shifted frequency, defined as $\Tilde{\omega} = m(\omega_{lm}-\Omega)$, where the pattern speed $\omega_{lm}$ is in general different from the planetary mean motion $n$ as per equation (\ref{eq:olm}). In line with previous work \citep{MirandaRafikov2019,MirandaRafikov2020}, we combine equations \eqref{eq:density_pert_eqn}--\eqref{eq:uphi_eqn} with the equation of state \eqref{eq:eos} in order to arrive at the master equation
\begin{align}
\label{eq:master_eqn}
     \frac{d^2}{dr^2}\delta h & +\left\{ \frac{d}{dr}\ln\left(\frac{r\Sigma}{D}\right)-\frac{1}{L_T} \right\} \frac{d}{dr}\delta h \nonumber \\
    & - \left\{ \frac{2m\Omega}{r\Tilde{\omega}}\left[\frac{1}{L_T}+\frac{d}{dr}\ln\left(\frac{\Sigma \Omega}{D}\right)\right] \right. \nonumber \\
    & \left.+\frac{1}{L_T}\frac{d}{dr}\ln\left(\frac{r\Sigma}{L_T D}\right)+\frac{m^2}{r^2}+\frac{D}{c_{\text{s,iso}}^2}\right\} \delta h = \nonumber \\
    & = -\frac{d^2 \Phi_{lm}}{dr^2} - \left[ \frac{d}{dr}\ln\left(\frac{r\Sigma}{D}\right)\right]\frac{d\Phi_{lm}}{dr} \nonumber \\
    & + \left\{\frac{2m\Omega}{r\Tilde{\omega}}\left[\frac{d}{dr}\ln\left(\frac{\Sigma \Omega}{D}\right)\right]+\frac{m^2}{r^2}\right\}\Phi_{lm},
\end{align}
in terms of the prognostic variable $\delta h = \delta P/\Sigma$, which is a useful pseudo-enthalpy variable often employed in the linear planet-disc studies. Here
\begin{equation}
\label{eq:detuning}
    D = \kappa^2-\Tilde{\omega}^2,
\end{equation}
is the detuning of the Doppler-shifted forcing frequency from the epicylic frequency $\kappa$, whilst
\begin{equation}
\label{eq:thermal_length}
    \frac{1}{L_T} = \frac{d\ln c_{s,\text{iso}}^2}{dr},
\end{equation}
sets the characteristic thermal length scale $L_T$. Note that knowing $\delta h$ we can reconstruct the other physical variables according to 
\begin{align}
    \label{eq:dsigma_dh}
    \delta\Sigma & = \frac{\Sigma \delta h}{c_{s,\text{iso}}^2} , \\
    \label{eq:dur_dh}
    \delta u_r &= \frac{i}{D} \left[ \left( \Tilde{\omega}\frac{\partial}{\partial r}-\frac{2m\Omega}{r}\right)(\delta h+\Phi_{lm})-\frac{\Tilde{\omega}}{L_T}\delta h\right] , \\
    \label{eq:duphi_dh}
    \delta u_\phi & = \frac{1}{D} \left[ \left( \frac{\kappa^2}{2\Omega}\frac{\partial}{\partial r}-\frac{m\tilde{\omega}}{r}\right)(\delta h+\Phi_{lm})-\frac{\kappa^2}{2\Omega L_T}\delta h \right].
\end{align}
Equations (\ref{eq:master_eqn})--(\ref{eq:thermal_length}) appear identical to the (locally isothermal) equations (17)--(21) of \citet{MirandaRafikov2020} except for the different definitions of $\tilde\omega$ and $D$ (now involving $\omega_{lm}$) as well as the potential components $\Phi_{lm}$ in the right hand side.

In the rest of this work we will focus on a globally isothermal setup (constant $c^2_\mathrm{s,iso}$, i.e. $q=0$) for which $1/L_T \rightarrow 0$. This further simplifies the equations \eqref{eq:master_eqn}, \eqref{eq:dur_dh} and \eqref{eq:duphi_dh} by eliminating a number of terms. Our focus on the globally isothermal EoS is motivated by the fact that the angular momentum flux of the freely propagating density waves is conserved for such EoS, as opposed to the more general locally isothermal EoS, see \citet{MirandaRafikov2019b}. Conservation of the angular momentum flux is very important to be able to predict the amplitude of the planetary density waves, which will be used in future analyses.

\section{Numerical procedure}
\label{section:numerical}

Now we will describe the details of the numerical procedures used in solving the master equation (\ref{eq:master_eqn}) and constructing spatial maps of the perturbation pattern. 

\subsection{Solving the master equation}
\label{subsection:solving_master}

The master equation \eqref{eq:master_eqn} is rather complicated (even for the globally isothermal EoS) and requires a numerical solution for each $(l,m)$ forcing component of the perturbing potential $\Phi_{lm}$ (determination of $\Phi_{lm}$ is described next in \S\ref{subsection:extracting_potential}). Since this problem is essentially the same as that solved by \cite{MirandaRafikov2019, MirandaRafikov2020} we exploit their numerical methodology with some careful extensions. Here we will briefly summarise the procedure, but the interested reader is referred to Appendix A of \cite{MirandaRafikov2019} and the original description laid out by \cite{KorycanskyPollack1993} for more details.

First, two linearly-independent homogeneous solutions, unforced by the planetary potential, are integrated outwards from the corotation radius $r_\mathrm{c}$ where the pattern speed $\omega_{lm}$ matches the background orbital flow rate $\Omega$, i.e. $\Omega(r_\mathrm{c})=\omega_{lm}$. Note that this position varies in general as the pattern speed changes for different $(l,m)$ combinations. Then an additional inhomogeneous solution, set by the $\Phi_{lm}$ forcing on the right hand side of equation \eqref{eq:master_eqn}, is integrated outwards as well. The true solution is some linear combination of these three results, which is determined by imposing the outgoing WKB wave boundary conditions as per \cite{TsangLai2008}. 

This initial solution is then made more robust in an iterative fashion as follows. We take the resulting values of $\delta h$ and $d \delta h/dr$ at the corotation location as new initial conditions for the forced differential equation. Integrating outwards then produces an updated inhomogeneous profile for $\delta h$. Once again the boundary conditions set the necessary homogeneous parts which are generally diminished by this process. This iterative procedure ensures that the resultant solution of the master equation \eqref{eq:master_eqn} is not overly sensitive to the initial (arbitrary) choice of initial conditions. 

As a final step, it is desirable to eliminate any spurious oscillations in the so called ``phase gradient error" from the solution near the boundaries \citep{KorycanskyPollack1993,MirandaRafikov2019}. This is a measure of the difference between the expected wave phase, set by the characteristic WKB wavenumber, and that found from the solution as d$\text{Arg}(\delta h)/dr$. Oscillations in this diagnostic are suggestive of unwanted incoming waves entering the domain. The amplitudes of these oscillations are reduced by using one step of the root finding Netwon-Raphson method, which adjusts the boundary condition prescription in order to minimise the phase gradient error function \citep{MirandaRafikov2019}. This, in turn, slightly updates the final solution. Although a small refinement, this step is important for proper handling of subtle interference effects when many modes are superimposed.

\subsection{Extracting the Fourier components of the perturbing potential}
\label{subsection:extracting_potential}

In order to solve the inhomogeneous version of the master equation \eqref{eq:master_eqn}, we must supply our solver with the radial structure of the $(l,m)$ component $\Phi_{lm}$ of the perturbing potential. In the zero eccentricity case considered previously this is simply done by computing the softened Laplace coefficients (e.g. see equation (13) of \cite{MirandaRafikov2019}). However, now that there are two intrinsic periodicities, we numerically extract each $\Phi_{lm}$ from equation \eqref{eq:fourier_expansion} via the double integral 
\begin{align}
    \label{eq:phi_mode_integral}
    \Phi_{lm}(r) &= \frac{1}{2\pi^2}\int_{t = 0}^{2\pi/n} \int_{\phi = 0}^{2\pi} \Phi(r,\phi,t) \cos(m\phi-l n t) \,d\phi\,dt \\
             &= -\frac{GM_{\text{p}}}{2\pi^2}\int_{t = 0}^{2\pi/n} \int_{\phi = 0}^{2\pi} \frac{\cos(m\phi-l n t) \,d\phi\,dt}{\left[r^2+r_{\text{p}}^2-2 r r_{\text{p}} \cos\psi+\epsilon^2\right]^{1/2}} , 
\end{align}
where $\psi(t) = \phi-\phi_{\text{p}}(t)$ denotes the azimuthal displacement from the planet and $\epsilon$ is the softening parameter employed to prevent singularities at the location of the planet when $r - r_{\text{p}} = \psi = 0$.

Calculation of this integral requires knowledge of the orbital trajectory of the planet $(r_{\text{p}}(t),\phi_{\text{p}}(t))$ over time. To get that, we solve\footnote{We employ the python package \texttt{kepler.py} which uses an algorithm based on the work of \cite{Markley1995} and \cite{Nijenhuis1991}.} Kepler's equation for the eccentric anomaly $E(t)$ as a function of the mean anomaly $M(t) = n t $, as given by equation (2.52) of \cite{MurrayMcDermott1999}.  This then provides the true anomaly of the planet such that
\begin{align}
\label{eq:phi_p}
    \cos\phi_{\text{p}}(t) = \frac{\cos E(t)-e}{1-e\cos E(t)}, ~~~~~~
    \sin\phi_{\text{p}}(t) = \frac{\sqrt{1-e^2}\sin E(t)}{1-e\cos E(t)},
\end{align}
and the radial position of the planet 
\begin{equation}
    \label{eq:r_p}
    r_{\text{p}}(t) = \frac{a(1-e^2)}{1+e \cos\phi_{\text{p}}(t)}.
\end{equation}
The double integral in equation \eqref{eq:phi_mode_integral} is then numerically computed using an equally weighted quadrature on a uniform grid with $( n_t , n_\phi )$ cells in $t$ and $\phi$ respectively. Note that $( n_t , n_\phi )$ must be chosen sufficiently high to avoid Nyquist-Shannon sampling artefacts for higher order modes. As we find in the convergence study performed in \S\ref{section:convergence}, we typically want the maximum mode number for $m$ to be over 100, which suggests taking $n_t$ and $n_\phi$ at least as large as 200.

The solver routine employed to integrate the master equation \eqref{eq:master_eqn} requires the $\Phi_{lm}$ to be calculated at each radial point as demanded by the adaptive step-size of its integration over $r$. Computing a double integral at each step is time-inefficient and so we resort to interpolation instead. We first calculate the $\Phi_{lm}$ on a logarithmically spaced grid defined between the inner radius $r_{\text{min}}$ and the outer radius $r_{\text{max}}$, at a set number of points, $n_r$. Then a cubic interpolation is used between these points for arbitrary values of $r$ to supply $\Phi_{lm}$ to the solver of the master equation.

\subsection{Mode superposition}

The quantum number $m$ controls the azimuthal structure of the modes. The previous numerical simulations of ZZ22 exhibit sharp, caustic features with significant information conveyed by the fine details of the perturbation pattern, suggesting that we should calculate a sufficiently high number of modes in $m$; we denote this number $m_{\text{max}}$. 

Meanwhile the $l$ quantum number encodes the time dependence of the pattern caused by the eccentricity of the planetary orbit. One can show that the coefficients of the modal expansion of the classical disturbing function are proportional to $e^{|l-m|}$ (see \cite{MurrayMcDermott1999}). Thus for small eccentricities, the ``off-diagonal" contributions are expected to quickly drop off as $l$ departs from $m$. We then set the range of $l$ that we consider about each $m$ according to the parameter $|l-m|_{\text{max}}$ --- the maximum value of $|l-m|$.

We need to be a little careful about which modes should be included in the net summation. Indeed, physically the density waves are launched from Linblad resonances where the Doppler shifted forcing frequency $\tilde{\omega}=m(\omega_{lm}-\Omega)$ matches the epicylic frequency $\kappa$. If we consider a disc with no pressure support, we can easily equate these frequencies and find the resonant locations to be
\begin{equation}
    \label{eq:linblad_resonances}
    \frac{r_{\pm}}{a} = \left(\frac{m \pm 1}{l}\right)^{2/3}, 
\end{equation}
where the plus and minus signs correspond to the inner and outer Linblad resonances respectively. Now, consider the following cases. When $m=1$ and $l \geq 1$ the inner Linblad resonance moves to $r_{-}\to  0$ and hence no resonance occurs in the inner part of the disc. However, there is still an outer Linblad resonance. In this case, the modal solution will be wave-like exterior to the Linblad resonance and evanescent interior to it. Despite the lack of an outgoing wave at the interior boundary, use of a higher order ODE solver is still able to find well behaved solutions which can be incorporated into the Fourier summation. Meanwhile, taking $m=1$ and $l<0$, would result in a non-physical resonance location and should be ignored.

For the case $m = 1$ and $l = 0$ the ratio given by equation \eqref{eq:linblad_resonances} for $r_{-}$ is indeterminate. In fact, this resonance corresponds to the specific condition that $\Omega(r) - \kappa(r) = 0$ which is satisfied everywhere in the disc for our assumed uniform pressure setup (see Section \ref{subsection:parameters}). This is an apsidal resonance wherein the precession frequency of disc material equals the precession frequency of the planet (which is zero in our Keplerian system). Such a secular resonance behaves distinctly different from a mean-motion Linblad resonance and will instead manifest itself as a global eccentric mode which depends sensitively on the boundary conditions. Such forced global eccentric modes have been studied by \cite{TeyssandierOgilvie2016a} and \cite{Lubow2022} but here we will neglect them since we are focusing on the launching of localised spiral features which can be treated as a separate effect. As such we will only consider mode contributions in the range $m \geq 1$ and $l \ge 1$ \footnote{In practice all the $m=1$ contributions are found to be subdominant and do not drastically affect the net morphology. Thus all these complications could be avoided in future by neglecting the $m=1$ Fourier terms. Note that this would also conveniently remove the indirect term from consideration.}.

\subsection{Parameters}
\label{subsection:parameters}

In this work we perform a range of targeted linear calculations to be compared with the numerical runs of ZZ22. For that reason, we adopt a setup for the disc analogous to their study, namely, a uniform background disc density with $p=0$. Furthermore, as stated earlier, we adopt a globally isothermal EoS with $q=0$ such that $L_T^{-1} \to 0$ and the wave angular momentum flux is conserved away from the wave excitation region. The disc aspect ratio at the semi-major axis of the planetary orbit is taken to be $h_{\text{p}} = 0.1$ and the gravitational softening length is set to be $\epsilon = 0.3 H$. We extract the planetary perturbing potential harmonics using the double quadrature method with $n_{\phi} = 1024$ and $n_t = 1024$. The disc extends between $r_{\text{min}} = 0.1 a$ and $r_{\text{max}} = 5 a$ and the interpolation scheme is used with $n_r = 1000$. The inhomogeneous refinement step (as described in \S\ref{subsection:solving_master}) is performed twice, whilst the phase gradient error correction is performed once. As motivated by the convergence study in \S\ref{section:convergence}, we adopt $m_{\text{max}} = 150$ and $|l-m|_{\text{order max}} = 8$ for the modal superposition. With this setup, we examine the morphology of the density wakes produced for three different values of planetary eccentricity $e = (0.1,0.25,0.5)$. 
\section{Results}
\label{section:results}

\subsection{Density wake morphology}
\label{section:eccentricity_results}

\begin{figure*}
    \centering
    \includegraphics[width=\textwidth]{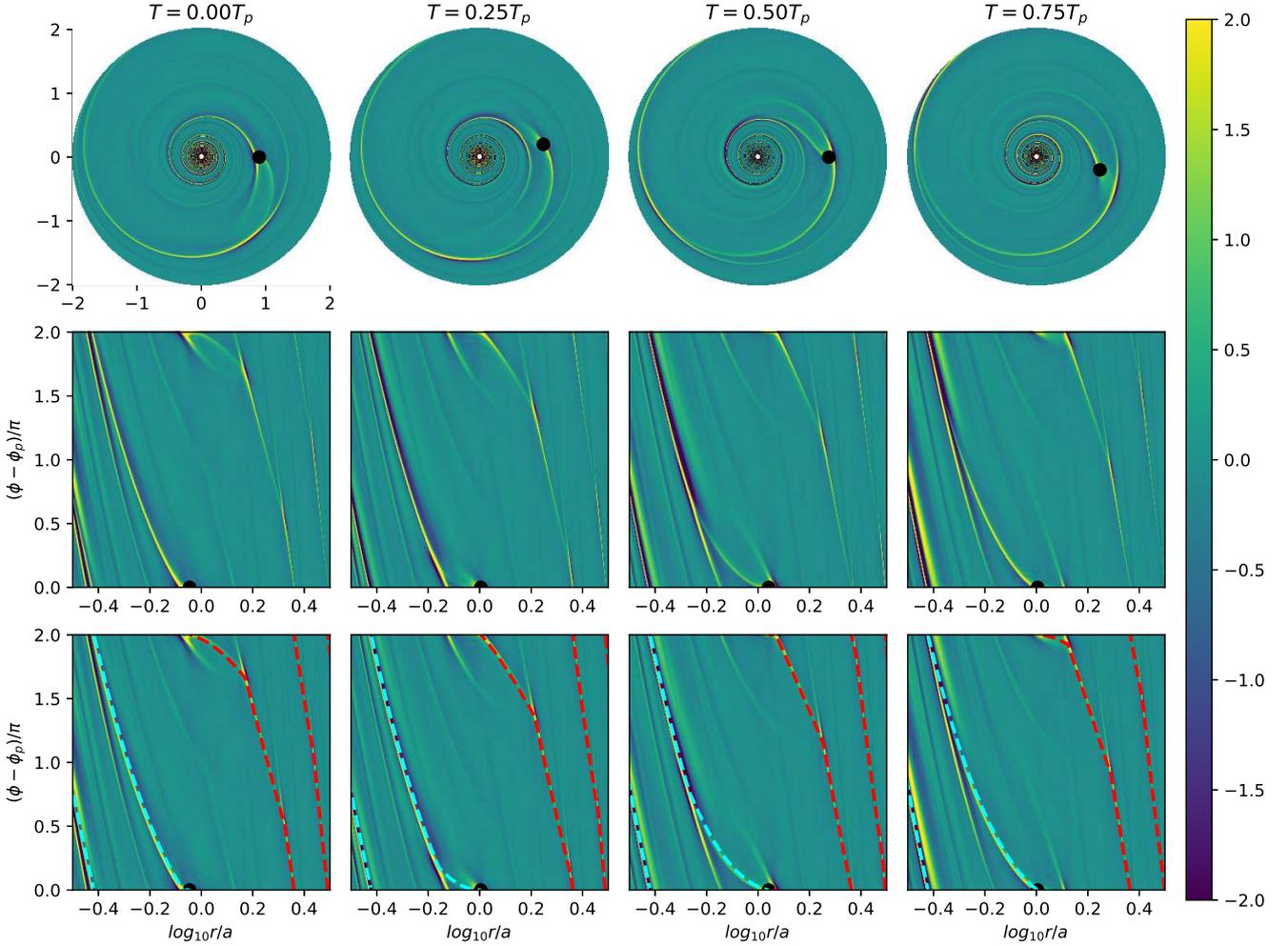}
    \caption{Scaled surface density perturbation $(M_{\text{p}} / M_{\text{th}})^{-1 }\delta\Sigma / \Sigma $ plotted for an eccentric planet with $e=0.1$. Each column shows snapshots at different moments of time (i.e. different planetary orbital phase) $(0.0,0.25,0.50) T_{\text{p}}$. A movie visualising the time-dependent evolution is available online. \textit{Top row}: A polar plot of the wake morphology in a frame corotating with the guiding centre of the planet. \textit{Middle row}: The spirals are unwrapped and shown in a Cartesian coordinate system, with planetary azimuthal location always aligned with zero (i.e. different from the top row). \textit{Bottom row}: The heuristic wavelet model of ZZ22 is over-plotted as dashed lines with the cyan/red denoting inwards/outwards propagating waves.}
    \label{fig:0.1_sigma_maps}
\end{figure*}

\begin{figure*}
    \centering
    \includegraphics[width=\textwidth]{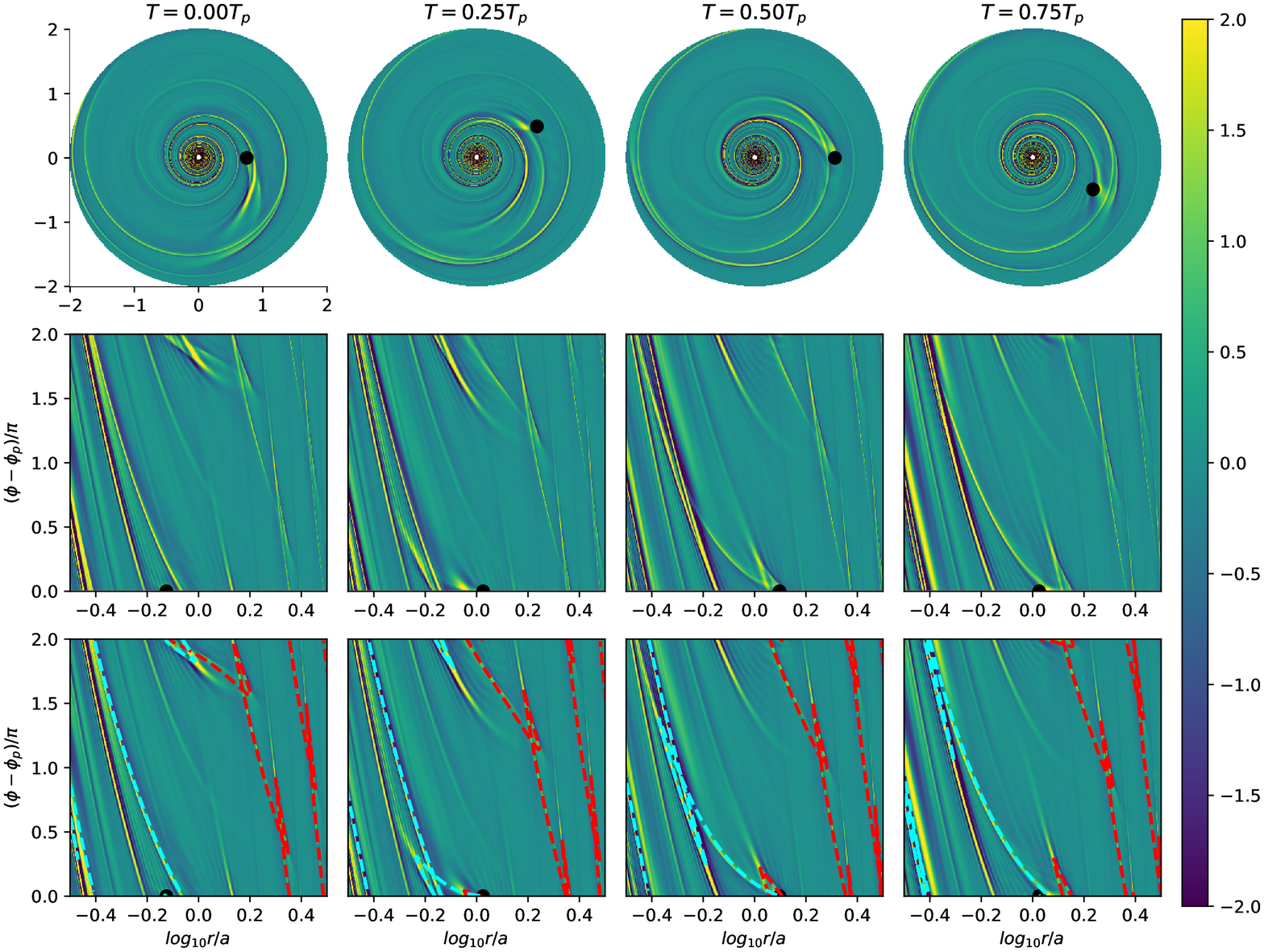}
    \caption{The same as Fig.~\ref{fig:0.1_sigma_maps} but for a planet with $e = 0.25$. A movie visualising the time-dependent evolution is available online.}
    \label{fig:0.25_sigma_maps}
\end{figure*}

\begin{figure*}
    \centering
    \includegraphics[width=\textwidth]{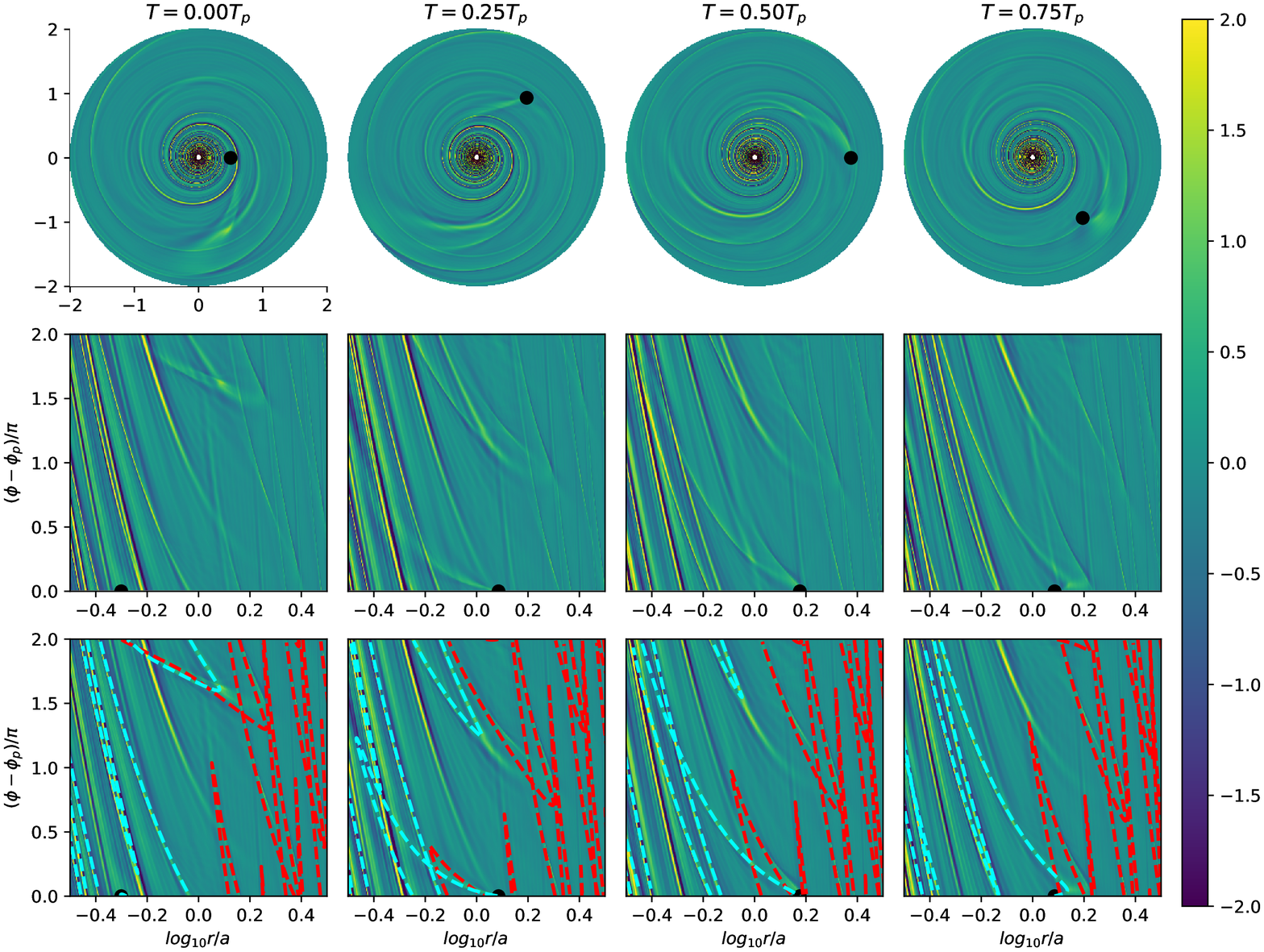}
    \caption{The same as Fig.~\ref{fig:0.1_sigma_maps} but for a planet with $e = 0.5$. A movie visualising the time-dependent evolution is available online.}
    \label{fig:0.5_sigma_maps}
\end{figure*}

Using our linear, fully time-dependent framework developed in previous sections, we now construct the surface density perturbation patterns for three different values of the eccentricity $e = (0.1,0.25,0.5)$ and compare them with the results of ZZ22. In each case we plot the surface density normalised by the background profile $\Sigma$ and scaled by the ratio of the planetary mass to the thermal mass $M_{\text{p}}/M_{\text{th}}$. The results are shown in Figures \ref{fig:0.1_sigma_maps}, \ref{fig:0.25_sigma_maps} and \ref{fig:0.5_sigma_maps} and animated movies showing the time evolution can be found in the online supplementary material. 

In each case, the upper row shows the snapshot of the (scaled) surface density perturbation $\delta\Sigma$ in polar coordinates in the frame co-rotating with the planetary guiding centre at 4 moments of time $T = (0.0,0.25,0.5,0.75)T_{\text{p}}$. The middle row unwraps the azimuthal dimension and plots the wake structure on a Cartesian grid, with the radial direction logarithmically scaled. Note that the azimuthal origin is translated so the planet always sits at $\phi = 0$. Finally in the bottom row we see the same figure but now overlaid with the wake locations predicted by the simplified theory of ZZ22. 

To compute the latter we launch wavelets from the planet location every $(1/800) T$ in time and track them according to the simple Huygens-like approach of ZZ22. In contrast to their method, we launch wavelets over 5 orbits so the wavelets have more time to propagate to the very edge of the domain, thus fully establishing the global pattern. Wavelet locations can thus be better compared with the time-dependent wake predicted by our theory at all phases. Connecting all these wavelets then traces out wake features. The inner propagating wavelets form the cyan dashed line, whilst the outer propagating wavelets form the red dashed line. For more details on this heuristic method we refer the reader to ZZ22.  

Our time-dependent linear theory is clearly capable of reproducing the numerical results found by ZZ22 who used a 2D implementation of the \texttt{FARGO} code \citep{Masset2000}. Indeed, direct comparison of our figures \ref{fig:0.1_sigma_maps}, \ref{fig:0.25_sigma_maps} and \ref{fig:0.5_sigma_maps} with their figures 5, 6 and 7 computed for $M_{\text{p}} = 3\times10^{-6}M_{*}=3\times10^{-3}M_\mathrm{th}$, shows excellent agreement. We are able to capture all the complex morphological features intrinsic to the eccentric orbits including the bifurcating spirals which exhibit cusp-like switch-backs termed `V points' by ZZ22 in the $e = (0.25,0.5)$ runs. These form due to the supersonic motion of the planet relative to the background gas as it `overtakes' the emitted wave-packets. This leads to a Mach cone-like structure behind the planet which is then sheared by the background flow. Here our calculations emphasise that this is an intrinsically linear feature of the interaction. 

As the eccentricity is increased the patterns become ever more complicated. The inner and outer propagating spiral arms sometimes cross and the planet can become detached from the wave. The over-plotted wavelet paths of ZZ22 show decent agreement with our linear theory. Indeed, akin to their comparisons with 2D numerical simulations, this approximate method is able to capture some of the main features of the linear wake. However, quite importantly, our linear theory is able to pick out a number of missing features and offer a far better correspondence with the numerical simulations. 

For example, in Fig.~\ref{fig:0.1_sigma_maps} at $T=(0.0-0.25) T_{\text{p}}$ we clearly see a bifurcating arm branching off the main outwards propagating spiral, which is not captured by the red dashed line. This appears to be launched as the planet moves clockwise and downwards through its apocentre on its epicylic path at $T = 0.50 T_{\text{p}}$, as seen in the upper row of Fig.\ref{fig:0.1_sigma_maps}. A similar feature can be seen for the inner region most clearly at $T=(0.5-0.75 T_{\text{p}})$ as an earlier spiral detaches from the cyan track. This is seen to be launched at $T = 0.0 T_{\text{p}}$ as the planet moves clockwise and upwards during its pericentre passage. 

In Figures \ref{fig:0.25_sigma_maps} and \ref{fig:0.5_sigma_maps} the pattern becomes more complex with many overlapping features. In fact, for the $e=0.5$ run, some of the inner spiral features appear to be external to the planetary location at $T = 0$ due to the large amplitude radial motion of the planet. While the method of ZZ22 picks out many of these features, it again fails to capture some of the elongated `V-point tails', which they attribute to the non-local nature and dispersion of the individual wavelets. Our method meanwhile reproduces all these features. 

Furthermore, our method provides us with information not only about the spatial structure of the perturbation but also about the {\it amplitude} of each mode forming the pattern. The latter is missing in the Huygens-like method of ZZ22 but is present in their direct numerical simulations. Although simulation results of ZZ22 do not explicitly show a colour bar, we see that their wakes becomes more diffuse and lower in amplitude as the planetary eccentricity increases, in agreement with the results of our semi-analytical calculations.

\subsection{Convergence study}
\label{section:convergence}

To ensure that the results produced by our semi-analytical framework are robust, we carried out a convergence study with respect to the number of modes that we use in reconstructing the perturbation pattern, i.e. the parameters $m_{\text{max}}$ and $|l-m|_{\text{max}}$. 

To do this in a quantitative fashion, we came up with the following convergence metric. Consider computing the perturbation pattern $\delta\Sigma_1$ for some given upper values of $m_{\text{max}}$ and $|l-m|_{\text{max}}$ across the full $(r,\phi)$ grid with $N$ points in total. This can be subtracted from the pattern $\delta\Sigma_2$ computed for some other $m_{\text{max}}$, $|l-m|_{\text{max}}$ to produce a difference map $\delta\Sigma_1-\delta\Sigma_2$. We introduce a {\it difference metric} $\Delta$ as 
\begin{equation}
\label{eq:difference_metric}
    \Delta(1,2) = \sum_{\phi} \sum_{r} \frac{(\delta\Sigma_1-\delta\Sigma_2)^2}{N}
\end{equation}
where the summation takes place over the output $(r,\phi)$-grid with $N$ points. The arguments of $\Delta$, denoted $(1,2)$, represent the two parameter sets $\{m_{\text{max},i}$, $|l-m|_{\text{max},i}\}$ for $i=1,2$. We now use $\Delta$ to look at the convergence in terms of $m_{\text{max}}$ and $|l-m|_{\text{max}}$ for the three eccentricity runs $e=(0.1, 0.25, 0.5)$ using the perturbation maps at orbital phase $T=0$ in each case. 

\begin{figure}
    \centering
    \includegraphics[width = \columnwidth]{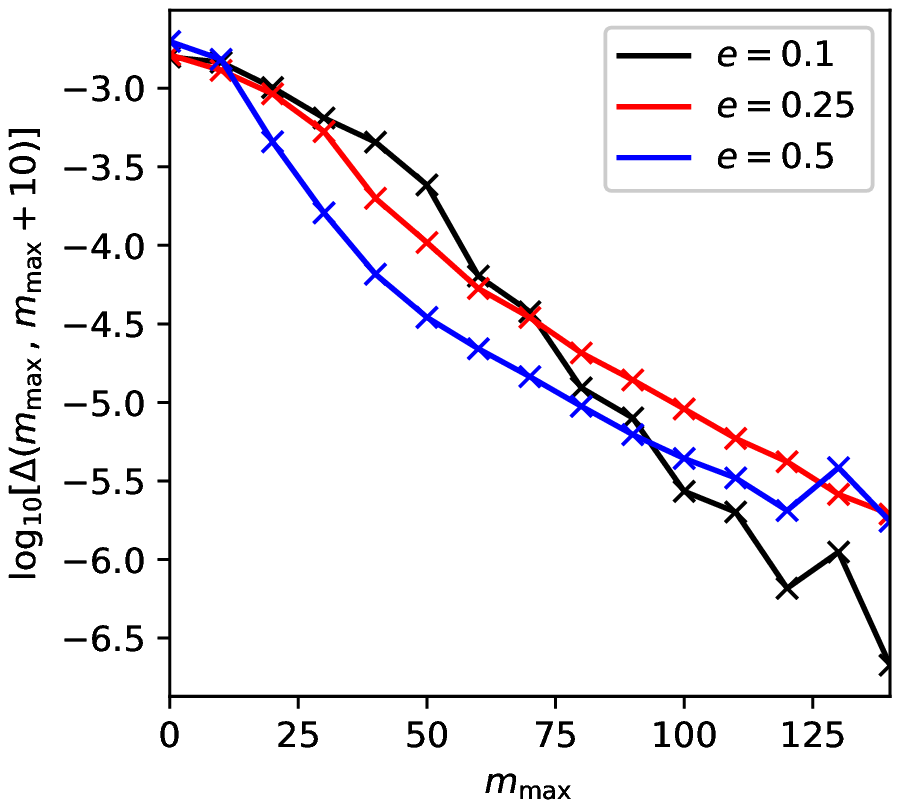}
    \caption{The convergence of our linear mode theory with increasing upper values of $m$ included in the summation. The y-axis plots the difference metric, given by equation \eqref{eq:difference_metric}, for two superpositions with upper values of $m_{\text{max}}$, as denoted on the x-axis, and $m_{\text{max}}+10$. The different colours (black, red, blue) denote the convergence properties for eccentricities (0.1, 0.25, 0.5) respectively. 
    \label{fig:convergence_m}}
\end{figure}
\begin{figure}
    \centering
    \includegraphics[width = \columnwidth]{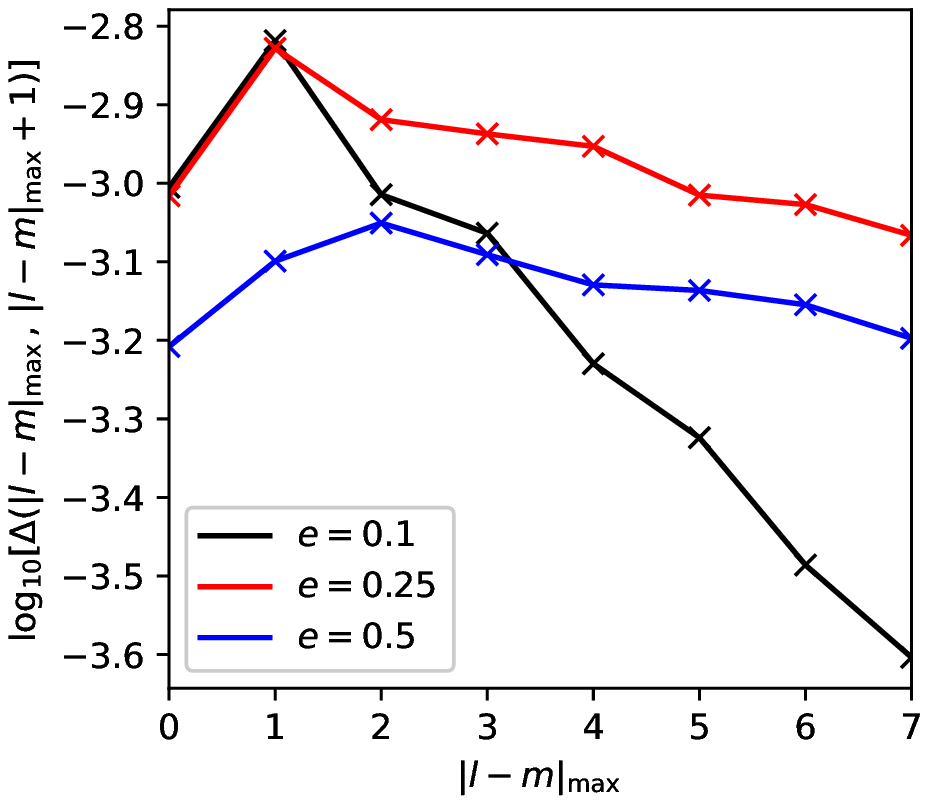}
    \caption{The convergence of our linear mode theory with increasing upper values of $|l-m|$ included in the summation. The y-axis plots the difference metric, given by equation \eqref{eq:difference_metric}, for two superpositions with upper values of $|l-m|_{\text{max}}$, as denoted on the x-axis, and $|l-m|_{\text{max}}+1$. The different colours (black, red, blue) denote the convergence properties for eccentricities (0.1, 0.25, 0.5) respectively.}
    \label{fig:convergence_e}
\end{figure}
In Fig.~\ref{fig:convergence_m} we examine the convergence in terms of $m_{\text{max}}$ by looking at the difference metric $\Delta(m_{\text{max}},m_{\text{max}}+10)$ between the two calculations with fixed $|l-m|_{\text{max}} = 8$ but $m_{\text{max}}$ differing by 10. The x-axis shows the value of $m_{\text{max},1}$ corresponding to the first calculation of the perturbation pattern, which uses fewer modes (the second calculation uses $m_{\text{max},2} = m_{\text{max},1}+10$). The vertical axis shows the logarithm of the difference metric $\Delta(m_{\text{max},1},m_{\text{max},2})$. The three different eccentricity runs are denoted by different colours.

One can see that for all values of $e$, the differences are largest for lower $m_{\text{max}}$ and it is these modes that most significantly contribute to the final converged image of the perturbation pattern. As $m_{\text{max}}$ is increased, $\Delta$ drops off in an approximately exponential fashion. By $m_{\text{max}}=80$ one finds $\Delta$ reduced by approximately 2 orders of magnitude compared to $m_{\text{max}}=10$. Thus, we can assume that the main features of the wake pattern are captured sufficiently well when $m_{\text{max}}=80-100$ is used. Notice that the slope of the drop off is slightly shallower for the red and blue lines corresponding to the higher eccentricity runs $e=(0.25,0.5)$. This might be expected due to the presence of sharp azimuthal structures associated with the wave bifurcations, cusps and crossings observed by ZZ22 (see \S\ref{section:eccentricity_results}).

In a similar vein, in Fig.~\ref{fig:convergence_e} we plot the difference metric when $m_\text{max} = 150$ is held fixed and instead $|l-m|_{\text{max}}$ differs by 1 between the two calculations of $\delta\Sigma$. Now the x-axis shows the value of $|l-m|_{\text{max},1}$ to be compared with $|l-m|_{\text{max},2} = |l-m|_{\text{max},1}+1$. The vertical axis shows the logarithm of $\Delta(|l-m|_{\text{max},1},|l-m|_{\text{max},2})$. Each planetary eccentricity is again shown as a different coloured line. 

For the smallest eccentricity, $e = 0.1$, the black line shows a clear drop off in $\Delta$ as we move towards higher $|l-m|_{\text{max}}$. In this case we see the dominant contributions are captured within $|l-m|_{\text{max}} \leq 4$ with an exponential drop off beyond this value. However, for the higher eccentricity runs $e = (0.25, 0.5)$ shown by the red and blue lines respectively, $\Delta$ exhibits a much broader profile with a shallower drop-off as one increases $|l-m|_{\text{max}}$. Indeed, this should be expected since the time dependent nature of the pattern becomes more important for larger eccentricities. 

Visual inspection of the surface density perturbation maps reveals that the main wake features are well captured by our method already for $|l-m|_{\text{max}}=4$ even at high planetary eccentricities, but they are slightly shifted and more diffuse compared with the perturbation maps computed for $|l-m|_{\text{max}}= 8$. Our perturbation metric (\ref{eq:difference_metric}) is not very well suited for capturing such changes, since even slight spatial shifts of narrow features with properly computed amplitudes would result in significant values of the difference metric $\Delta$, despite the key features being sufficiently resolved (certainly up to observational standards).   

To summarise, these findings suggest that our scheme has converged suitably well for  $m_{\text{max}} = 150$ and $|l-m|_{\text{max}} = 8$, reassuring us in the robustness of the results shown in \S\ref{section:eccentricity_results}.

\section{Discussion}
\label{section:discussion}

Our results clearly demonstrate the applicability of the semi-analytical linear framework developed in this work for understanding the morphology of the density waves driven by eccentric planets in protoplanetary discs. While the fundamental concepts of eccentric planet-disc interaction have been around since the seminal study of \citet{GoldreichTremaine1980}, our work is the first one that uses the relevant mathematical tools to predict the spatial perturbation patterns excited by the eccentric planets in the linear regime.

Our calculations also reaffirm the results of direct numerical simulations performed by ZZ22 in the limit of small planet masses. Sharp, shock-like features found by ZZ22 are usually thought of as owing their existence to the nonlinear effects naturally present in simulations (shocks due to the finite-amplitude effects). Our calculations show that such features emerge in linear theory and are well captured by our linear formalism. 

Our semi-analytical method has several advantages compared with previous approaches. First, it can be more numerically efficient\footnote{Using $m_{\text{max}} = 150$ and $|l-m|_{\text{max}} = 8$ requires ~24 wall-clock hours on a $2\times18$ core \texttt{Skylake} $6140$ $2.3 GHz$ processor.} than full numerical simulations depending on the number of modes used for reconstruction of the perturbation pattern. Of course, the higher order modes become much more oscillatory and hence present a more expensive computational challenge when solving the master equation \eqref{eq:master_eqn}. However, the convergence study shown in figures \ref{fig:convergence_m} and \ref{fig:convergence_e} has shown that in many cases (particularly for lower values of $e$) the dominant wake features are well captured for $m \leq 80$ and $|l-m| \leq 4$. Taking fewer modes can offer a considerable speedup in computation time. 

Furthermore, the global nature of our linear solutions means that the final wake pattern is established ab initio at any moment of time (i.e. planetary orbital phase). In comparison, when running numerical simulations one needs to be careful about introducing the planetary perturbation in the disc and giving the disturbance enough time to fully propagate through the computational domain before the (quasi-)steady state becomes fully established.

Our approach also gives us quantitative information on the amplitude of each mode. This can help us isolate the modes responsible for each pattern speed $\omega_{lm}$ and systematically examine how the separate spiral features manifest themselves. Indeed, previous authors have shown using linear theory that the constructive interference of individual modes leads to a well-defined primary spiral arm near the planet \citep{OgilvieLubow2002,Rafikov2002}, whilst the dispersive effects lead to secondary and tertiary arms forming further from the planet \citep{MirandaRafikov2019}. The inherent time-dependence of our problem complicates the details of such mode superposition and can lead to the appearance of multiple spiral features, even next to the planet, owing to its epicyclic motion. This issue merits further attention in future studies. Furthermore, our modal approach can isolate the torque density and angular momentum flux contributions due to each individual harmonic, which are key for informing the back reaction onto planet migration and eccentricity evolution rates -- something that will also be explored in future work.

Knowledge of the wave amplitude is also important for assessing the role of nonlinear effects for wake evolution. Our linear approach is strictly valid and provides excellent approximation to the numerical results in the limit of $M_\mathrm{p}\ll M_\mathrm{th}$, although we expect it to work well (at least at the qualitative level) even for $M_\mathrm{p}\sim M_\mathrm{th}$, based on the experience gained for planets on circular orbits \citep{Goodman2001,Rafikov2002,Dong2011,Cimer2021}. Compared to the circular case, splitting of the density wake into a number of spiral arms is likely to reduce the amplitude of each individual arm, delaying its nonlinear evolution and shocking. On the other hand, some of the arms are sharper (feature higher gradients) than the characteristic one-armed spiral in the $e=0$ case, which could speed up the non-linear evolution of these particular wake features. This issue merits further study, but in any case, we expect the shocking length obtained in  \citet{Goodman2001} and \citet{Rafikov2002} to remain a reasonable predictor of the radial scale over which the nonlinear effects could become important.

Our study has established a proof-of-concept agreement with the numerical results of ZZ22 for the globally isothermal and uniform surface density disc. However, future studies should also perform a more general exploration of the physical parameter space. The mathematical machinery developed in \S\ref{section:framework} can be readily extended to include more detailed thermodynamics including cooling prescriptions, such as the ones explored by \citet{MirandaRafikov2020,MirandaRafikov2020b} in the case of zero eccentricity planet-disc interactions. Cooling was found to have a significant effect on angular momentum flux transport and wake structure in these studies, which will undoubtedly impact the eccentric regime as well. Moreover, despite the focus and motivation of this paper, our method is not only applicable to eccentric planet-disc interaction, but can also be used for any time-dependent perturbing potential. 

\section{Conclusions}
\label{section:conclusions}

We developed a versatile linear semi-analytical framework to explore the spiral wave morphology produced by an eccentric planet embedded in a two-dimensional gaseous disc. The inherent time-dependence of the eccentric planetary orbit leads to a Fourier summation for the perturbation over two quantum numbers governing the azimuthal and temporal periodicities of the wake pattern. We use this theory to compute the shape of the density wake for a globally isothermal and uniform surface density disc for three values of the planetary eccentricity $e = (0.1,0.25,0.5)$. Our results show excellent agreement with the numerical simulations of \cite{ZhuZhang2022} for the same set of parameters and are able to capture all of the key features produced by eccentric planets including multiple spiral features with different pattern speeds, bifurcating spiral arms, V-points and planet detachment from the wake structure. We also compare our synthesised images of the density perturbation with the heuristic wavelet method of \cite{ZhuZhang2022}. We find reasonable agreement with the main wake features, however, our full linear method is able to better reproduce the complexity of the resultant perturbation features. This proof-of-concept work demonstrates the power of linear theory in probing the eccentric planet-disc interactions. Future work will further explore the physical parameter space of the problem including the different underlying disc profiles, disc cooling prescriptions, and so on.  

\section*{Acknowledgements}
The authors would like to thank the anonymous reviewer for their insightful comments and suggestions. 
CWF would also like to acknowledge helpful conversations with Gordon Ogilvie and Nicolas Cimerman. This research was supported by an STFC studentship (CWF), as well as the STFC grant ST/T00049X/1 and Ambrose Monell Foundation (RRR). 

\section*{Data Availability}
 
Data used in this paper is available from the authors upon reasonable
request.

\typeout{}
\bibliographystyle{mnras}
\bibliography{main} 




\bsp	
\label{lastpage}
\end{document}